\documentclass[a4paper]{article}

\usepackage[margin=1in]{geometry} 

\usepackage[T1]{fontenc}
\newcommand{\changefont}[3]{
\fontfamily{#1} \fontseries{#2} \fontshape{#3} \selectfont}

\changefont{ptm}{m}{n}

\usepackage{setspace} 
\usepackage{graphicx}

\usepackage{amsfonts}
\usepackage{amsmath}
\usepackage{amssymb}
\usepackage{graphics}
\usepackage{mathrsfs}
\usepackage{color}

\newtheorem{theorem}{Theorem}[section]

\newtheorem{corollary}{Corollary}[section]
\newtheorem{lemma}{Lemma}[section]

\newtheorem{definition}{Definition}[section]

\long\def\symbolfootnote[#1]#2{\begingroup%
\def\thefootnote{\fnsymbol{footnote}}\footnote[#1]{#2}\endgroup} 

\begin{document}

\begin{center}
\Large \textbf{Unpredictable Solutions of Quasilinear Systems with Discontinuous Right-Hand Sides}
\end{center}

\begin{center}
\normalsize \textbf{Mehmet Onur Fen$^{1,}\symbolfootnote[1]{Corresponding Author. E-mail: monur.fen@gmail.com}$, Fatma Tokmak Fen$^2$} \\
\vspace{0.2cm}
\textit{\textbf{$^1$Department of Mathematics, TED University, 06420 Ankara, Turkey}} \\

\vspace{0.1cm}
\textit{\textbf{$^2$Department of Mathematics, Gazi University, 06500 Ankara, Turkey}} \\
\vspace{0.1cm}
\end{center}

\vspace{0.3cm}

\begin{center}
\textbf{Abstract} 
\end{center}

\vspace{-0.2cm}

\noindent\ignorespaces It is rigorously proved under certain assumptions that a quasilinear system with discontinuous right-hand side possesses a unique unpredictable solution. The discontinuous perturbation function on the right-hand side is defined by means of an unpredictable sequence. A Gronwall-Coppel type inequality is utilized to achieve the main result, and the stability of the unpredictable solution is discussed. Examples with exponentially asymptotically stable and unstable unpredictable solutions are provided.

\vspace{0.2cm}
 
\noindent\ignorespaces \textbf{Keywords:} Unpredictable solution, Unpredictable sequence, Quasilinear system, Poincar\'e chaos, Discontinuous Right-Hand Side

\vspace{0.2cm}

\noindent\ignorespaces \textbf{Mathematics Subject Classification:} 34A34, 34A36, 34C15, 34D20

\vspace{0.6cm}

\section{Introduction} \label{intro}
 
A new type of chaos, named Poincar\'e chaos, was introduced in paper \cite{Fen16}. The main feature of this type of chaos is its initiation through a single trajectory in the dynamics. An unpredictable trajectory is capable of generating Poincar\'e chaos, and such a trajectory is necessarily positively Poisson stable. Sensitive dependence on initial conditions is a prominent feature of the corresponding quasi-minimal set \cite{Fen16}.
 
The notions of Poincar\'e chaos and unpredictable trajectory were developed in papers \cite{Fen17a}-\cite{Fen18}. Unpredictable solutions of differential equations was first taken into account in study \cite{Fen17a} by means of the Bebutov dynamics \cite{Sell71}. A continuous unpredictable function was constructed in paper \cite{Fen17} through the symbolic dynamics and logistic map.  Additionally, the paper \cite{Fen18} was devoted to the existence and uniqueness of unpredictable solutions in systems with delay.  
The papers \cite{Miller19}-\cite{Thakur21}, on the other hand, are concerned with Poincar\'e chaos in topological spaces. The concept of unpredictable point was generalized to semiflows with arbitrary acting topological monoids by Miller \cite{Miller19}. Additionally, Thakur and Das \cite{Thakur20} studied Poincar\'e chaos on the product of semiflows. Moreover, the presence of an unpredictable point in an hyperspatial semiflow was demonstrated in \cite{Thakur21}.

In this paper we take into account the system
\begin{eqnarray} \label{mainsystem}
x'(t) = Ax(t) + f(t,x(t)) + g(t),
\end{eqnarray}
where the eigenvalues of the constant matrix $A \in \mathbb R^{ p \times p}$ have nonzero real parts and $f : \mathbb R \times \mathbb R^p \to \mathbb R^p$ is a continuous function which is $\omega$-periodic in the first variable, i.e., there is a positive number $\omega$ such that $$f(t+ \omega, x) = f(t,x)$$ for every $t \in \mathbb R$ and $x \in \mathbb R^p$. The piecewise constant function $g : \mathbb R \to \mathbb R^p$ is defined through the equation
\begin{eqnarray} \label{discontineqn}
g(t) = \sigma_k
\end{eqnarray}
for $t \in [\theta_{k}, \theta_{k+1})$, where $\left\{\sigma_k\right\}_{k \in \mathbb Z}$ is a bounded sequence in $\mathbb R^p$ with $\displaystyle\sup_{k \in \mathbb Z}  \left\|\sigma_k \right\| \leq M_{\sigma}$ for some positive number $M_{\sigma}$ and $\left\{\theta_k\right\}_{k \in \mathbb Z}$ is a sequence in $\mathbb R$ satisfying
\begin{eqnarray} \label{thetak}
\theta_{k+1} = \theta_k+ \omega
\end{eqnarray}
for each $k\in \mathbb Z$.

The main novelty of the present study is the usage of a discontinuous perturbation for quasilinear systems to generate an unpredictable solution. The reader is referred to the book \cite{Filippov88} for general information about systems with discontinuous right-hand sides. In this paper, we rigorously prove the existence and uniqueness of an unpredictable solution in the dynamics of system (\ref{mainsystem}) in the case that the sequence $\left\{\sigma_k\right\}_{k \in \mathbb Z}$ is unpredictable. The asymptotic stability of the unpredictable solution is also discussed. The main source for the presence of an unpredictable solution is the piecewise constant perturbation function $g(t)$ used in (\ref{mainsystem}). Our results reveal that a discontinuous perturbation can give rise to the formation of an unpredictable solution in quasilinear systems.

Unpredictable solutions of systems with hyperbolic linear part were also investigated in paper \cite{Fen19}. There are two main differences of the present study compared to \cite{Fen19}. Firstly, in this paper the main source of the unpredictable solution is a discontinuous function defined through an unpredictable sequence, whereas a continuous perturbation is utilized in \cite{Fen19}. Secondly, we make use of a Gronwall-Coppel type inequality to prove the main result. However, the results of paper \cite{Fen19} are based on the contraction mapping principle.  

The rest of the paper is organized as follows. The transformation of system (\ref{mainsystem}) to a one with a block diagonal matrix of coefficients is discussed in the next section. Moreover, the existence and uniqueness of a bounded solution is investigated. It is proved in Section \ref{sec3} that system (\ref{mainsystem}) admits a unique unpredictable solution under the assumption that the Lipschitz constant of the nonlinear term is sufficiently small. The exponentially asymptotic stability of the unpredictable solution is also discussed in that section. Two examples which support the theoretical results are provided in Section \ref{examplessec}. The main difference of these examples is that in the first one the unpredictable solution is unstable, whereas the second one comprises a $3-$dimensional exponentially asymptotically stable unpredictable trajectory. Finally, some concluding remarks are given in Section \ref{secconc}.

\section{Preliminaries} \label{sec_prelim}
 
In the remaining parts of the paper we use the Euclidean norm for vectors and the spectral norm for square matrices. 
 
In system (\ref{mainsystem}) we assume that $\Re e \lambda_i>0$ for $i=1,2,\ldots,r$ and $\Re e \lambda_i<0$ for $i=r+1,r+2,\ldots,p$ with $1\leq r<p,$ where $\lambda_i,$ $i=1,2,\ldots,p,$ are the eigenvalues of the matrix $A$ and $\Re e \lambda_i$ denotes the real part of $\lambda_i.$

There exists a nonsingular matrix $P$ such that $$P^{-1} A P = \textrm{diag} \left(A_+, A_-\right),$$ where $A_+$ is an $r \times r$ matrix whose eigenvalues have positive real parts and  $A_-$ is a $(p-r) \times (p-r)$ matrix whose eigenvalues have negative real parts  \cite{Hale80}. Moreover, there exist numbers $N \geq 1$ and $\gamma>0$ such that $\left\|e^{A_+ t}\right\| \leq N e^{\gamma t}$ for $t \leq 0$ and $\left\|e^{A_- t}\right\| \leq N e^{-\gamma t}$ for $t \geq 0$  \cite{Hale80}.
  
 The following assumptions are required.
  \begin{itemize}
  \item[\textbf{(A1)}] There exists a positive number $M_f$ such that $\displaystyle \sup_{(t,x) \in [0, \omega] \times \mathbb R^p} \left\|f(t,x)\right\|  \leq M_f;$
 \item[\textbf{(A2)}] There exists a positive number $L_f$ such that $ \left\|f(t,x_1) -f(t,x_2)\right\|  \leq L_f \left\|x_1-x_2\right\|$ for each $x_1,x_2 \in \mathbb R$.
\end{itemize}

Under the substitution $x(t)=Py(t)$, system (\ref{mainsystem}) is transformed to 
 \begin{eqnarray} \label{mainsystem2}
y'(t) = P^{-1} A P y(t) + P^{-1}f(t,Py(t)) + P^{-1} g(t).
\end{eqnarray}
Let us denote $$P^{-1}f(t,z) = \left(f_+(t, z),f_-(t,z)\right)^T $$ 
and 
$$P^{-1} g(t) = \left(g_+(t), g_-(t)\right)^T,$$
where $f_+(t, z)$, $g_+(t)$ are vector-valued functions with dimension $r$ and  $f_-(t, z)$, $g_-(t)$ are vector-valued functions with dimension $p-r$. 

One can show using the results of the book \cite{Hale80} that a function $$\phi(t)= \left(\phi_+(t),\phi_-(t)\right)^T$$ which is bounded on the whole real axis is a solution of system (\ref{mainsystem2}) if and only if the functions $\phi_+: \mathbb R \to \mathbb R^r$ and $\phi_-: \mathbb R \to \mathbb R^{p-r}$ respectively satisfy the relations
\begin{eqnarray} \label{phiplus}
\phi_+(t) = \displaystyle - \int_{t}^{\infty} e^{A_+(t-s)} \left( f_+(s,P \phi(s)) + g_+(s) \right) ds
\end{eqnarray}
and
\begin{eqnarray} \label{phiminus}
\phi_-(t) = \displaystyle \int^{t}_{-\infty} e^{A_-(t-s)} \left( f_-(s,P \phi(s)) + g_-(s) \right) ds.
\end{eqnarray} 
 
\begin{lemma} \label{boundedsolns}
If the assumptions $(A1)$, $(A2)$ are valid and $\sqrt{2} N L_{f} \left\|P\right\| \left\|P^{-1}\right\| - \gamma <0$, then system (\ref{mainsystem2}) possesses a unique solution $\phi(t)$ which bounded on the whole real axis such that 
\begin{eqnarray} \label{boundedineq1}
\displaystyle \sup_{t \in \mathbb R}  \left\|\phi(t)\right\| \leq \sqrt{2}N \left\|P^{-1}\right\|\gamma^{-1}(M_f +M_{\sigma}).
\end{eqnarray}  
\end{lemma} 
 
\noindent {\bf Proof.} Let $\mathcal{S}$ be the set of continuous and uniformly bounded functions $\phi: \mathbb R \to \mathbb R^p$ satisfying the inequality $\left\|\phi\right\|_{\infty} \leq \sqrt{2}N\left\|P^{-1}\right\| \gamma^{-1} (M_f +M_{\sigma})$, where $\left\|\phi\right\|_{\infty} =\displaystyle \sup_{t \in \mathbb R} \left\|\phi(t)\right\|$.
Define an operator $\Omega$ on the set $\mathcal{S}$ by the equation $\Omega \phi (t) = \left(\left(\Omega \phi\right)_+ (t), \left(\Omega \phi\right)_-(t) \right)^T$, where
\begin{eqnarray*}
\left(\Omega \phi\right)_+ (t) = \displaystyle - \int_{t}^{\infty} e^{A_+(t-s)} \left( f_+(s,P \phi(s)) + g_+(s) \right) ds
\end{eqnarray*}
and
\begin{eqnarray*}
\left(\Omega \phi\right)_- (t) = \displaystyle \int^{t}_{-\infty} e^{A_-(t-s)} \left( f_-(s,P \phi(s)) + g_-(s) \right) ds.
\end{eqnarray*} 
Making use of the estimations
\begin{eqnarray*}
\left\|\left(\Omega \phi\right)_+ (t)\right\| & \leq &  \displaystyle   \int_{t}^{\infty} \big\|e^{A_+(t-s)}\big\|  \left\| f_+(s,P \phi(s)) + g_+(s)  \right\| ds \\
&\leq & \displaystyle   \int_{t}^{\infty} N \left\|P^{-1}\right\| \left(M_{f} + M_{\sigma}\right) e^{\gamma (t-s)}   ds \\
&=& N\left\|P^{-1}\right\|\gamma^{-1}(M_f +M_{\sigma})
\end{eqnarray*}
and 
\begin{eqnarray*}
\left\|\left(\Omega \phi\right)_- (t)\right\| & \leq &  \displaystyle   \int^{t}_{-\infty} \big\|e^{A_-(t-s)}\big\|  \left\| f_-(s,P \phi(s)) + g_-(s)  \right\| ds \\
&\leq & \displaystyle   \int^{t}_{-\infty} N \left\|P^{-1}\right\| \left(M_{f} + M_{\sigma}\right) e^{-\gamma (t-s)}  ds \\
&=& N \left\|P^{-1}\right\| \gamma^{-1}(M_f +M_{\sigma}),
\end{eqnarray*}
we obtain that $\left\|\Omega\phi\right\|_{\infty} \leq \sqrt{2}N\left\|P^{-1}\right\|\gamma^{-1}(M_f +M_{\sigma})$. Thus, $\Omega  \left(\mathcal{S} \right) \subseteq \mathcal{S}.$

Now, let $\phi_1(t)$ and $\phi_2(t)$ be functions in $\mathcal{S}$. Then, we have 
\begin{eqnarray*}
\left\|\left(\Omega \phi_1\right)_+ (t)-\left(\Omega \phi_2\right)_+ (t)\right\| 
& \leq & \displaystyle \int_{t}^{\infty} \big\|e^{A_+(t-s)}\big\|  \left\|f_+(s,P \phi_1(s)) -f_+(s,P \phi_2(s))  \right\| ds \\
& \leq & \displaystyle \int_{t}^{\infty} N L_{f} \left\|P\right\| \left\|P^{-1}\right\| e^{\gamma (t-s)}  \left\| \phi_1(s) - \phi_2(s) \right\| ds \\
& \leq & N L_{f} \left\|P\right\| \left\|P^{-1}\right\| \gamma^{-1}  \left\| \phi_1 - \phi_2 \right\|_{\infty}
\end{eqnarray*}
and
\begin{eqnarray*}
\left\|\left(\Omega \phi_1\right)_- (t)-\left(\Omega \phi_2\right)_- (t)\right\| 
& \leq & \displaystyle \int^{t}_{-\infty} \big\|e^{A_-(t-s)}\big\| \left\|f_-(s,P \phi_1(s)) -f_-(s,P \phi_2(s))  \right\| ds \\
& \leq & \displaystyle \int^{t}_{-\infty} N L_{f} \left\|P\right\|  \left\|P^{-1}\right\| e^{-\gamma (t-s)}  \left\| \phi_1(s) - \phi_2(s) \right\| ds \\
& \leq & N L_{f} \left\|P\right\|  \left\|P^{-1}\right\| \gamma^{-1}   \left\| \phi_1 - \phi_2 \right\|_{\infty}.
\end{eqnarray*}
Therefore,
$$
\left\|\left(\Omega \phi_1\right)-\left(\Omega \phi_2\right) \right\|_{\infty} \leq \displaystyle \sqrt{2} N L_{f} \left\|P\right\| \left\|P^{-1}\right\| \gamma^{-1}  \left\| \phi_1 - \phi_2 \right\|_{\infty}.
$$
The operator $\Omega$ is contractive in accordance with the inequality $\sqrt{2} N L_{f} \left\|P\right\| \left\|P^{-1}\right\| - \gamma <0$. Hence, there exists a unique bounded solution $\phi(t)$ of (\ref{mainsystem2}) satisfying (\ref{boundedineq1}). $\square$

 In what follows we will denote 
 \begin{eqnarray} \label{nummpsi}
M_{\psi}= \sqrt{2}N\left\|P\right\|\left\|P^{-1}\right\|\gamma^{-1}(M_f +M_{\sigma}).
 \end{eqnarray}
A corollary of Lemma \ref{boundedsolns} is as follows.
 
\begin{corollary} \label{boundedsolnscor}
If the assumptions $(A1)$, $(A2)$ are valid and $\sqrt{2} N L_{f} \left\|P\right\| \left\|P^{-1}\right\| - \gamma <0$, then system (\ref{mainsystem}) possesses a unique solution $\psi(t)$ which bounded on the whole real axis such that 
$
\displaystyle \sup_{t \in \mathbb R}  \left\|\psi(t)\right\| \leq M_{\psi}.
$ 
\end{corollary}  
\noindent {\bf Proof.} Because the system (\ref{mainsystem2}) possesses a unique bounded solution $\phi(t)$ such that (\ref{boundedineq1}) holds by Lemma \ref{boundedsolns}, system (\ref{mainsystem}) admits the unique bounded solution $\psi(t) = P \phi(t)$ and the inequality $\displaystyle \sup_{t \in \mathbb R}  \left\|\psi(t)\right\| \leq M_{\psi}$ is valid. $\square$

\section{Existence of an Unpredictable Solution} \label{sec3}

The definitions of an unpredictable sequence and a uniformly continuous unpredictable function are as follows.

\begin{definition} (\cite{Fen18}) \label{unpseqquasi}
A bounded sequence $\left\{\sigma_k\right\}_{k \in \mathbb Z}$ is called unpredictable if there exist a positive number $\delta_{0}$ (the unpredictability constant) and sequences $\left\{\zeta_n\right\}_{n\in\mathbb N},$ $\left\{\eta_n\right\}_{n\in\mathbb N}$ of positive integers both of which diverge to infinity such that $\left\|\sigma_{k+\zeta_n} - \sigma_k \right\|\to 0$ as $n \to \infty$ for each $k$ in bounded intervals of integers and $ \left\|\sigma_{\zeta_n + \eta_n} - \sigma_{\eta_n}\right\| \geq \delta_{0}$ for each $n\in\mathbb N.$
\end{definition}

\begin{definition} (\cite{Fen18}) \label{unpfuncquasi}
A uniformly continuous and bounded function $\psi: \mathbb R \to \mathbb R^p$ is called unpredictable if there exist positive numbers $\epsilon_0$ (the unpredictability constant), $r$ and sequences $\left\{\mu_n\right\}_{n\in\mathbb N}$ and $\left\{\nu_n\right\}_{n\in\mathbb N}$ both of which diverge to infinity such that $\left\|\psi(t+\mu_n)-\psi(t)\right\| \to 0$ as $n \to \infty$ uniformly on compact subsets of $\mathbb R$ and $\left\|\psi(t+\mu_n)-\psi(t)\right\| \geq \epsilon_0$ for each $t \in [\nu_n-r,\nu_n+r]$ and $n \in \mathbb N.$
\end{definition}

The proof of the following Gronwall-Coppel type inequality is provided in paper \cite{Akhmet18}.

\begin{lemma} [\cite{Bainov92,Akhmet18}] \label{lemmagronwall}
Let $a$, $b$, $c$, and $d$ be constants such that $a \geq 0$, $b \geq 0$, $c>0$, $d>0$, and suppose that $\varphi(t) \in C\left([r_1,r_2], \mathbb R\right)$ is a nonnegative function satisfying
$$
\varphi(t) \leq a+b \left(e^{-d(t-r_1)} + e^{-d(r_2-t)}\right) + c \displaystyle \int_{r_1}^{r_2} e^{-d|t-s|} \varphi(s) ds, \ t\in [r_1,r_2].
$$
If $2c<d$, then 
$$
\varphi(t) \leq \displaystyle \frac{ad}{d-2c} + \frac{b}{c}(d-h)  \left(e^{-h(t-r_1)} + e^{-h(r_2-t)}\right) 
$$
for any $t\in[r_1,r_2]$, where $h=\sqrt{d^2-2cd}$.
\end{lemma}

Lemma \ref{lemmagronwall} is utilized in the proof of the next assertion. The following assumption is also needed.
\begin{itemize}
\item[\textbf{(A3)}] $2 N L_{f} \left\|P\right\| \left\|P^{-1}\right\| - \gamma <0.$
\end{itemize}

\begin{lemma} \label{requiredlemma}
Suppose that there is a sequence $\left\{\zeta_n\right\}_{n\in\mathbb N}$ of positive integers which diverges to infinity such that $\left\|\sigma_{k+\zeta_n} - \sigma_{k} \right\|\to 0$ as $n \to \infty$ for each $k$ in bounded intervals of integers. Then, under the assumptions $(A1)-(A3)$, the bounded solution $\psi(t)$ of system (\ref{mainsystem}) has the property that $\left\|\psi(t+\omega \zeta_n)-\psi(t)\right\| \to 0$ as $n \to \infty$ uniformly on compact subsets of $\mathbb R$.
\end{lemma}

\noindent \textbf{Proof.} Let $\epsilon$ be a fixed positive number and $\mathcal{C}$ be a compact subset of $\mathbb R$. There exist integers $\alpha$ and $\beta$ with $\alpha < \beta$ such that $\mathcal{C} \subseteq [\theta_{\alpha}, \theta_{\beta}]$.
Suppose that $\rho_0$ is a positive number satisfying
\begin{eqnarray} \label{lemmaineq1} 
\rho_0 < \displaystyle \frac{\gamma - 2NL_f \left\|P\right\| \left\|P^{-1}\right\|}{4N\left\|P^{-1}\right\|}
\end{eqnarray}
and $j$ is a natural number such that
\begin{eqnarray} \label{lemmaineq2} 
j > \displaystyle \frac{1}{\omega \gamma_0} \ln\left( \frac{8(M_f + M_{\sigma}) (\gamma - \gamma_0)}{\gamma L_f \left\|P\right\| \epsilon}\right),
\end{eqnarray}
where 
\begin{eqnarray} \label{lemmaeq3}
\gamma_0 = \big(\gamma^2 - 2\gamma N L_f  \left\|P\right\| \left\|P^{-1}\right\| \big)^{1/2}.
\end{eqnarray}

Because $\left\|\sigma_{k+\zeta_n}-\sigma_k\right\| \to 0$ as $n \to \infty$ for each $k$ in bounded intervals of integers, there exists a number $n_0 \in \mathbb N$ such that for $n \geq n_0$ we have $\left\|\sigma_{k+\zeta_n}-\sigma_k\right\| < \rho_0 \epsilon$ for each $k = \alpha - j, \alpha - j +1, \ldots, \beta +j-1 $. For that reason 
$
\left\|g(t + \omega \zeta_n) - g(t)\right\| < \rho_0 \epsilon
$
for $\theta_{\alpha -j} \leq t < \theta_{\beta+j}$.

In the remaining part of the proof we will show that $\left\|\phi(t+\omega \zeta_n)-\phi(t)\right\| \to 0$ as $n \to \infty$ uniformly on $\mathcal{C}$, where $\phi(t)= \left(\phi_+(t),\phi_-(t)\right)^T$ is the unique solution of system (\ref{mainsystem2}) which is bounded on the real axis.

Let $n \geq n_0$ be a fixed natural number. Using (\ref{phiplus}) and (\ref{phiminus}) one can respectively obtain that
\begin{eqnarray*}
\phi_+(t + \omega \zeta_n) - \phi_+(t)=
- \displaystyle \int_t^{\infty} e^{A_+ (t-s)} \big(  f_+(s, P\phi(s+\omega \zeta_n))  +g_+(s+\omega\zeta_n)- f_+(s, P\phi(s)) - g_+(s) \big) ds
\end{eqnarray*}
and
\begin{eqnarray*}
\phi_-(t + \omega \zeta_n) - \phi_-(t)=
\displaystyle \int^t_{-\infty} e^{A_- (t-s)} \big(  f_-(s, P\phi(s+\omega \zeta_n)) + g_-(s+\omega\zeta_n)- f_-(s, P\phi(s)) - g_-(s) \big) ds.
\end{eqnarray*}
These equations yield for $\theta_{\alpha-j} \leq t \leq \theta_{\beta+j}$ that
\begin{eqnarray*}
\left\|\phi_+(t + \omega \zeta_n) - \phi_+(t)\right\|
& <& N  \left\|P^{-1}\right\| \gamma^{-1} \rho_0 \epsilon \left(1-e^{-\gamma (\theta_{\beta+j}-t)}\right) \\
& + & 2N \left\|P^{-1}\right\|  \gamma^{-1} (M_f + M_{\sigma}) e^{-\gamma (\theta_{\beta+j}-t)} \\
&+& N L_f \left\|P\right\| \left\|P^{-1}\right\| \displaystyle \int_t^{\theta_{\beta+j}} e^{\gamma (t-s)} \left\|\phi(s + \omega \zeta_n) - \phi(s)\right\| ds
\end{eqnarray*}
and
\begin{eqnarray*}
\left\|\phi_-(t + \omega \zeta_n) - \phi_-(t)\right\|
& < & N  \left\|P^{-1}\right\| \gamma^{-1} \rho_0 \epsilon \left(1-e^{-\gamma (t-\theta_{\alpha-j})}\right) \\
&+ & 2N \left\|P^{-1}\right\|  \gamma^{-1} (M_f + M_{\sigma}) e^{-\gamma (t-\theta_{\alpha-j})} \\
&+& N L_f \left\|P\right\| \left\|P^{-1}\right\| \displaystyle \int^t_{\theta_{\alpha-j}} e^{-\gamma (t-s)} \left\|\phi(s + \omega \zeta_n) - \phi(s)\right\| ds.
\end{eqnarray*}
Therefore, 
\begin{eqnarray*}
\left\|\phi(t + \omega \zeta_n) - \phi(t)\right\| 
&<& 2 N  \left\|P^{-1}\right\| \gamma^{-1} \rho_0 \epsilon \\
& + &  2N \left\|P^{-1}\right\|  \gamma^{-1} (M_f + M_{\sigma}) \left(e^{-\gamma (t-\theta_{\alpha-j})} + e^{-\gamma (\theta_{\beta+j}-t)}\right) \\
&+& N L_f \left\|P\right\| \left\|P^{-1}\right\| \displaystyle \int^{\theta_{\beta+j}}_{\theta_{\alpha-j}} e^{-\gamma  \left|t-s\right| } \left\|\phi(s + \omega \zeta_n) - \phi(s)\right\| ds.
\end{eqnarray*}

One can attain by Lemma \ref{lemmagronwall} that the inequality
\begin{eqnarray*}
\left\|\phi(t + \omega \zeta_n) - \phi(t)\right\| 
\leq \displaystyle \frac{2N\left\|P^{-1}\right\| \rho_0 \epsilon}{\gamma-2 N L_{f} \left\|P\right\| \left\|P^{-1}\right\|}
+ \displaystyle \frac{2(M_f + M_{\sigma})}{\gamma L_f \left\|P\right\|} (\gamma - \gamma_0) \left(e^{-\gamma_0 (t-\theta_{\alpha-j})} + e^{-\gamma_0 (\theta_{\beta+j}-t)}\right)
\end{eqnarray*}
is valid for $\theta_{\alpha-j} \leq t \leq \theta_{\beta+j}$, where the number $\gamma_0$ is defined by (\ref{lemmaeq3}).
Owing to (\ref{lemmaineq1}) we have  $$\displaystyle \frac{2N\left\|P^{-1}\right\| \rho_0 \epsilon}{\gamma-2 N L_{f} \left\|P\right\| \left\|P^{-1}\right\|} < \frac{\epsilon}{2}.$$ Moreover, since (\ref{lemmaineq2}) holds, one can confirm for $\theta_{\alpha} \leq t \leq \theta_{\beta}$ that 
\begin{eqnarray*}
  \displaystyle \frac{2(M_f + M_{\sigma})}{\gamma L_f \left\|P\right\|} (\gamma - \gamma_0) \left(e^{-\gamma_0 (t-\theta_{\alpha-j})} +  e^{-\gamma_0 (\theta_{\beta+j}-t)}\right)  
  \leq    \displaystyle \frac{4(M_f + M_{\sigma})}{\gamma L_f \left\|P\right\|} (\gamma - \gamma_0) e^{-\gamma_0 j \omega}  
  <   \displaystyle \frac{\epsilon}{2}.
\end{eqnarray*}
Thus, 
$$
\left\|\phi(t + \omega \zeta_n) - \phi(t)\right\| < \epsilon, \ \ \theta_{\alpha} \leq t \leq \theta_{\beta}. 
$$
Accordingly $\left\|\phi(t+\omega \zeta_n)-\phi(t)\right\| \to 0$ as $n \to \infty$ uniformly on $\mathcal{C}$.

Since $\mathcal{C}$ is an arbitrary compact subset of $\mathbb R$, the inequality $$\left\|\psi(t+\omega \zeta_n)-\psi(t)\right\| \leq \left\|P\right\| \left\|\phi(t+\omega \zeta_n)-\phi(t)\right\|$$ implies that $\left\|\psi(t+\omega \zeta_n)-\psi(t)\right\| \to 0$ as $n \to \infty$ uniformly on compact subsets of $\mathbb R$. $\square$

The existence and uniqueness of an unpredictable solution of system (\ref{mainsystem}) is mentioned in the following theorem.

\begin{theorem} \label{hypmainthm1}
Suppose that the assumptions $(A1)-(A3)$ hold. If the sequence $\left\{\sigma_k\right\}_{k \in \mathbb Z}$ is unpredictable, then system (\ref{mainsystem}) possesses a unique unpredictable solution. Moreover, if all eigenvalues of the matrix $A$ have negative real parts, then the unpredictable solution is exponentially asymptotically stable.
\end{theorem}

\noindent \textbf{Proof.} System (\ref{mainsystem}) admits a unique solution $\psi(t)$ which is bounded on $\mathbb R$ such that
$\displaystyle \sup_{t \in \mathbb R}  \left\|\psi(t)\right\| \leq M_{\psi}$ according to Corollary \ref{boundedsolnscor}, where the number $M_{\psi}$ is defined by (\ref{nummpsi}). In the proof, we will show that the bounded solution $\psi(t)$ is unpredictable.

Since $\psi'(t) = A \psi(t) + f(t, \psi(t)) + g(t)$, the inequality
$$\displaystyle \sup_{t \in \mathbb R} \left\|\psi'(t)\right\|\leq  \left\|A\right\| M_{\psi} + M_{f} + M_{\sigma}$$
is valid. Hence, $\psi(t)$ is uniformly continuous.

In accordance with Definition \ref{unpseqquasi}, there exist a positive number $\delta_{0}$ and sequences $\left\{\zeta_n\right\}_{n\in\mathbb N},$ $\left\{\eta_n\right\}_{n\in\mathbb N}$ of positive integers both of which diverge to infinity such that $\left\|\sigma_{k+\zeta_n} - \sigma_k \right\|\to 0$ as $n \to \infty$ for each $k$ in bounded intervals of integers and 
\begin{eqnarray} \label{sigmaineq1}
 \left\|\sigma_{\zeta_n + \eta_n} - \sigma_{\eta_n}\right\| \geq \delta_{0}
\end{eqnarray}
 for each $n\in\mathbb N.$

Lemma \ref{requiredlemma} implies that $\left\|\psi(t+\mu_{n})-\psi(t)\right\| \to 0$ as $n \to \infty$ uniformly on compact subsets of $\mathbb R$, where $\mu_n = \omega \zeta_n$ for each $n \in \mathbb N$. The sequence $\left\{\mu_n\right\}_{n \in \mathbb N}$ diverges to infinity since the same feature is true for the sequence $\left\{\zeta_n\right\}_{n\in\mathbb N}$. Therefore, in order to show that $\psi(t)$ is unpredictable it is enough to verify the existence positive numbers $\epsilon_0$, $r$ and a sequence $\left\{\nu_n\right\}_{n \in \mathbb N}$, which diverges to infinity, such that $\left\|\psi(t+\mu_n)-\psi(t)\right\| \geq \epsilon_0$ for each $t \in [\nu_n-r,\nu_n+r]$ and $n \in \mathbb N$.

For each natural number $k$, let us denote the terms of the sequence $\left\{ \sigma_k\right\}_{k \in \mathbb N} $ by $\sigma_k = \left(\sigma_k^1, \sigma_k^2, \ldots, \sigma_k^p\right)^T$, where $\sigma_k^i \in \mathbb R$ for each integer $i$ with $1 \leq i \leq p$. 

Fix a natural number $n$. According to inequality (\ref{sigmaineq1}), there exists an integer $i_0$  with $1 \leq i_0 \leq p$ such that
\begin{eqnarray*}  
 \big|\sigma^{i_0}_{\zeta_n + \eta_n} - \sigma^{i_0}_{\eta_n}\big| \geq \displaystyle \frac{\delta_{0}}{\sqrt{p}}. 
\end{eqnarray*}
It can be obtained using equation (\ref{discontineqn}) that $g(t) = \sigma_{\eta_n}$ and $g(t+\mu_n) = \sigma_{\zeta_n+\eta_n}$ for $\theta_{\eta_n} \leq t < \theta_{\eta_n +1}$.
Therefore,
\begin{eqnarray*}   
\Big\| \displaystyle \int_{\theta_{\eta_n}}^{\theta_{\eta_n+1}} \left( g(s+\mu_n) -g(s)  \right) ds \Big\| 
= \Big\| \displaystyle \int_{\theta_{\eta_n}}^{\theta_{\eta_n+1}} \left(  \sigma_{\zeta_n+\eta_n} - \sigma_{\eta_n}  \right) ds \Big\|
\geq \omega   \big|\sigma^{i_0}_{\zeta_n + \eta_n} - \sigma^{i_0}_{\eta_n}\big| \geq \displaystyle \frac{\omega \delta_{0}}{\sqrt{p}}.
\end{eqnarray*} 
Using the equations
$$
\psi(\theta_{\eta_n +1}) = \psi(\theta_{\eta_n}) + \displaystyle \int_{\theta_{\eta_n}}^{\theta_{\eta_n +1}} \left( A \psi(s) + f(s, \psi(s)) + g(s) \right) ds
$$ 
and
$$
\psi(\theta_{\zeta_n+\eta_n +1}) = \psi(\theta_{\zeta_n+\eta_n}) + \displaystyle \int_{\theta_{\eta_n}}^{\theta_{\eta_n +1}} \left( A \psi(s+\mu_n) + f(s, \psi(s+\mu_n)) + g(s+\mu_n) \right) ds,
$$ 
one can attain that
\begin{eqnarray*}   
\left\| \psi(\theta_{\zeta_n+\eta_n +1}) - \psi(\theta_{\eta_n +1})  \right\| 
& \geq & \Big\| \displaystyle \int_{\theta_{\eta_n}}^{\theta_{\eta_n+1}} \left( g(s+\mu_n) -g(s)  \right) ds \Big\| 
- \left\| \psi(\theta_{\zeta_n+\eta_n}) - \psi(\theta_{\eta_n})  \right\|  \\
&& -  \Big\| \displaystyle \int_{\theta_{\eta_n}}^{\theta_{\eta_n+1}} A \left( \psi(s+ \mu_n) - \psi(s) \right)  ds  \Big\| \\
&& -  \Big\| \displaystyle \int_{\theta_{\eta_n}}^{\theta_{\eta_n+1}} \left( f(s,\psi(s+ \mu_n)) - f(s,\psi(s)) \right)  ds  \Big\| \\
& \geq & \displaystyle \frac{\omega \delta_0}{\sqrt{p}} - \left\| \psi(\theta_{\zeta_n+\eta_n}) - \psi(\theta_{\eta_n})  \right\| \\
&& - \omega \left(\left\|A\right\| + L_f\right) \sup_{t \in [\theta_{\eta_n}, \theta_{\eta_n+1}]} \left\| \psi(t+ \mu_n) - \psi(t) \right\|.
\end{eqnarray*} 
Hence, we have 
\begin{eqnarray*}   
  \sup_{t \in [\theta_{\eta_n}, \theta_{\eta_n+1}]} \left\| \psi(t+ \mu_n) - \psi(t) \right\| \geq \displaystyle \frac{\omega \delta_0}{\sqrt{p} [2+ \omega \left(\left\|A\right\| + L_f\right) ]}.
\end{eqnarray*} 
Now, let $\nu_n$ be a point in the interval $[\theta_{\eta_n}, \theta_{\eta_n+1}]$  such that
 $$\sup_{t \in [\theta_{\eta_n}, \theta_{\eta_n+1}]} \left\| \psi(t+ \mu_n) - \psi(t) \right\| = \left\|\psi(\nu_n + \mu_n) - \psi(\nu_n)\right\|,$$
and define the number
$$r=  \frac{\omega \delta_0}{4 \sqrt{p} \left[M_{\psi} (\left\|A\right\|+L_f) +M_{\sigma}\right]   \left[2+ \omega (\left\|A\right\|+L_f) \right]}.$$ 
The equation
\begin{eqnarray*}
\psi(t+\mu_n) - \psi(t) & = & \psi(\nu_n+\mu_n) - \psi(\nu_n) + \displaystyle \int_{\nu_n}^t A \left( \psi(s+\mu_n) - \psi(s) \right) ds \\
&& + \displaystyle \int_{\nu_n}^t \left( f(s,\psi(s+\mu_n)) - f(s, \psi(s))   \right) ds \\
&& + \displaystyle \int_{\nu_n}^t \left(g(s+ \mu_n) - g(s) \right) ds
\end{eqnarray*} 
implies for $t \in [\nu_n-r, \nu_n+r]$ that
\begin{eqnarray*}
\left\|\psi(t+\mu_n) - \psi(t)\right\| & \geq & \left\| \psi(\nu_n+\mu_n) - \psi(\nu_n) \right\| 
 - \Big|  \displaystyle \int_{\nu_n}^t \left\|A\right\| \left\|\psi(s+\mu_n) - \psi(s) \right\| ds \Big| \\
 && - \Big|  \displaystyle \int_{\nu_n}^t \left\|  f(s,\psi(s+\mu_n)) - f(s, \psi(s))  \right\|  ds \Big| 
 - \Big|  \displaystyle \int_{\nu_n}^t \left\|  g( s+\mu_n ) - g( s )  \right\|  ds \Big| \\
 & \geq & \frac{\omega \delta_0}{\sqrt{p} \left[2+ \omega (\left\|A\right\|+L_f) \right]}
 - 2r [M_{\psi}(\left\|A\right\|+L_f) + M_{\sigma}].
\end{eqnarray*}  
Therefore, $\left\|\psi(t+\mu_n) - \psi(t)\right\| \geq \epsilon_0$ for each $t \in [\nu_n-r, \nu_n+r]$ and $n \in \mathbb N$, where
 $$\epsilon_0 = \displaystyle \frac{\omega \delta_0}{2 \sqrt{p} \left[2+ \omega (\left\|A\right\|+L_f) \right]}.$$ Additionally, $\nu_n \to \infty$ as $n \to \infty$ since $ \eta_n \to \infty$ as $n \to \infty$. Thus, the bounded solution $\psi(t)$ of system (\ref{mainsystem}) is unpredictable. 
 
In the case that all eigenvalues of the matrix $A$ have negative real parts the exponentially asymptotic stability of the unpredictable solution can be proved in a very similar way to the proof of Theorem 5.4 \cite{Corduneanu77}. $\square$

\section{Examples} \label{examplessec}

Two examples which support the result of Theorem \ref{hypmainthm1} are presented in this section. In the first one, we consider a system in which the matrix of coefficients of the linear part admits both positive and negative eigenvalues. The second example provides us the opportunity to simulate the unpredictable behavior due to the asymptotic stability. 

It was shown in paper \cite{Fen17} that  the logistic map 
\begin{eqnarray} \label{logisticmap}
z_{k+1} = \lambda z_k (1-z_k),
\end{eqnarray}
where $k \in \mathbb Z$, admits an unpredictable orbit  for $3+(2/3)^{1/2} \leq \lambda \leq 4$. Moreover, for those values of $\lambda$  the unit interval $[0,1]$ is invariant under the iterations of (\ref{logisticmap}) \cite{Hale91}. In both of the examples, an unpredictable orbit of the logistic map (\ref{logisticmap}) is utilized.

\subsection{Example 1}  \label{examplesec1}
 
Let us fix an unpredictable orbit $\left\{z^*_k\right\}_{k \in \mathbb Z}$ of the logistic map (\ref{logisticmap}) with $\lambda=3.93$, whose terms belong to the interval $[0,1]$. One can confirm using Theorem $3.2$ \cite{Fen18} that $\left\{\sigma_k\right\}_{k\in\mathbb Z}$, where $$\sigma_k= \left((0.6+z^*_k)^2, 1.4 z^*_k+ \sin(z^*_k)\right)^T, \ k\in \mathbb Z,$$ is an unpredictable sequence.
 
We take into account the system
\begin{eqnarray} \label{discontsyst1}
\begin{array}{l}
x_1'(t)=4x_1(t)+x_2(t) + 0.4\cos t + g_1(t)\\
x_2'(t)=x_1(t)-2x_2(t) + 0.3 \arctan(x_1(t)) + g_2(t),
\end{array}
\end{eqnarray} 
where the functions $g_i: \mathbb R \to \mathbb R$, $i=1, 2$, are defined by $g_1(t)= (0.6+z^*_k)^2$ and $g_2(t)= 1.4 z^*_k+ \sin(z^*_k)$ for $t \in [\theta_k, \theta_{k+1})$, $k \in \mathbb Z$, and $\left\{\theta_k\right\}_{k \in \mathbb Z}$ is the sequence satisfying $\theta_0=0.1$ such that the equation (\ref{thetak}) holds with $\omega=2 \pi$. 
 
System (\ref{discontsyst1}) is in the form of (\ref{mainsystem}), where
$$A = 
\begin{pmatrix} 
4 & 1 \\ 
1 & -2 
\end{pmatrix},$$
$$f(t,x_1,x_2)=(0.4\cos t, 0.3 \arctan(x_1) )^T,$$
and 
$$g(t)=(g_1(t),g_2(t))^T.$$
It can be verified that
$P^{-1} A P = \textrm{diag} \left(1+\sqrt{10}, 1-\sqrt{10}\right),$ where
$$P = 
\begin{pmatrix} 
 1 & \sqrt{10}-3 \\ 
\sqrt{10}-3 & -1 
\end{pmatrix}.$$
The conditions $(A1)-(A3)$ hold for system (\ref{discontsyst1}) with $N=1$, $\gamma=\sqrt{10}-1$, $M_{\sigma}=3.5091$, $M_f =0.6182$, and $L_f=0.3$. Therefore, according to Theorem \ref{hypmainthm1}, system (\ref{discontsyst1}) possesses a unique unpredictable solution.

Due to the presence of a positive eigenvalue of the matrix $A$, the unpredictable solution of system (\ref{discontsyst1}) is unstable. For that reason it is not possible to visualize the unpredictable behavior of the system by simulations. In the next example, we provide a system which possesses an exponentially asymptotically stable unpredictable solution.

\subsection{Example 2}  \label{examplesec2}

In this example we denote by $\left\{z^*_k\right\}_{k \in \mathbb Z}$ an unpredictable orbit of the logistic map (\ref{logisticmap}) with $\lambda=3.91$, whose terms belong to the interval $[0,1]$,
and consider the system
\begin{eqnarray} \label{discontsyst2}
\begin{array}{l}
x_1'(t)=-5x_1(t)-2.5 x_2(t) + 0.1 \cos (x_3(t)) + g_1(t) \\
x_2'(t)=2x_1(t)-3x_2(t) + 0.5 \sin (4t) + g_2(t) \\
x_3'(t)=-1.5x_3(t) + 0.2 \tanh(x_1(t)) + g_3(t),
\end{array}
\end{eqnarray} 
where the functions $g_i : \mathbb R \to \mathbb R$, $i=1,2,3$, are defined by the equations $g_1(t)= 4.5 z^*_k$, $g_2(t)= 3.7 z^*_k$, and $g_3(t)= 3.9 z^*_k$ for $t \in [\theta_k, \theta_{k+1})$, $k \in \mathbb Z$, the sequence $\left\{\theta_k\right\}_{k \in \mathbb Z}$ satisfies equation (\ref{thetak}) with $\omega=\pi/2$, and $\theta_0=0.3$. It is worth noting that  $\left\{\sigma_k\right\}_{k\in\mathbb Z}$ with $\sigma_k= \left(4.5 z^*_k, 3.7 z^*_k, 3.9 z^*_k\right)^T$, $k\in \mathbb Z$, is an unpredictable sequence according to Theorem $3.2$ \cite{Fen18}.

System (\ref{discontsyst2}) is in the form of (\ref{mainsystem}), where
$$A = 
\begin{pmatrix} 
-5 & -2.5 & 0\\ 
2 & -3 & 0 \\
0 & 0 & -1.5
\end{pmatrix},$$
$$f(t, x_1, x_2, x_3) = (0.1 \cos(x_3), 0.5 \sin (4t), 0.2 \tanh(x_1))^T,$$ 
and
$$g(t)=(g_1(t), g_2(t), g_3(t))^T.$$
The eigenvalues of the matrix $A$ are $-4+2i$, $-4-2i$, and $-1.5$. The equation $A=QBQ^{-1}$ is valid, where
$$B = 
\begin{pmatrix} 
-4 & -2 & 0\\ 
2 & -4 & 0 \\
0 & 0 & -1.5
\end{pmatrix}, \ \
Q = 
\begin{pmatrix} 
 2 & -1 & 0\\ 
0 & 2 & 0 \\
0 & 0 & 1
\end{pmatrix}.$$
Thus,
$$e^{At} = Q \begin{pmatrix} 
e^{-4t} \cos(2t) & -e^{-4t} \sin(2t) & 0\\ 
e^{-4t} \sin(2t) & e^{-4t} \cos(2t) & 0 \\
0 & 0 & e^{-1.5 t}
\end{pmatrix} Q^{-1} $$
and the inequality $\left\|e^{At}\right\| \leq Ne^{-\gamma t}$, $t \geq 0$, is satisfied for $$N= \left\|Q\right\| \left\|Q^{-1}\right\|=\displaystyle \left((9 + \sqrt{17})/2\right)^{1/2}$$ and $\gamma =1.5$. 

The conditions of Theorem \ref{hypmainthm1} hold for system (\ref{discontsyst2}) with $M_{\sigma}=7.0108$, $M_f =0.5478$, and $L_f=0.2$. Therefore, system (\ref{discontsyst2}) admits a unique unpredictable solution. Moreover, the unpredictable solution of (\ref{discontsyst2}) is exponentially asymptotically stable since the eigenvalues of the matrix $A$ have negative real parts. 

\begin{figure}[ht!] 
\centering
\includegraphics[width=11.45cm]{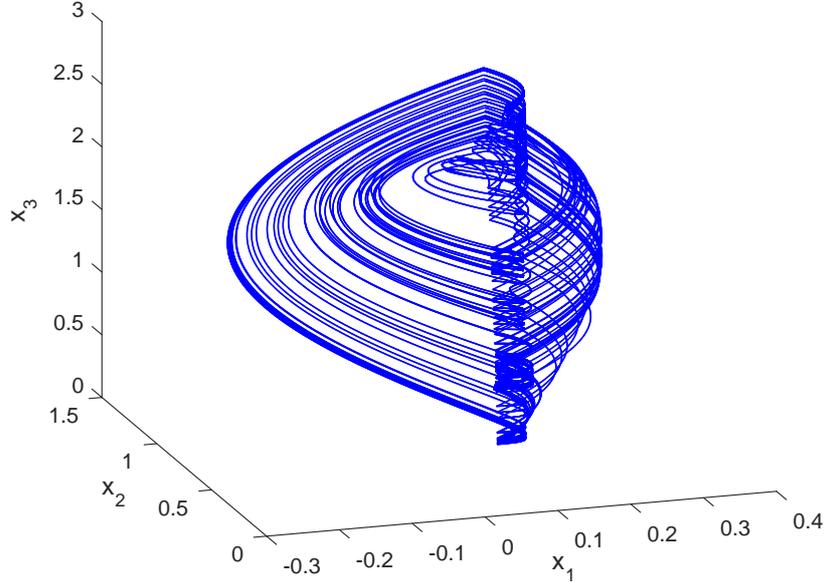}
\caption{$3-$dimensional trajectory of system (\ref{discontsyst3}) corresponding to the initial data $x_1(0.3)= 0.12$, $x_2(0.3)= 0.35$, and $x_3(0.3)= 0.83$. The figure reveals the existence of an unpredictable solution.}
\label{quasifig1}
\end{figure}

\begin{figure}[ht] 
\centering
\includegraphics[width=15.00cm]{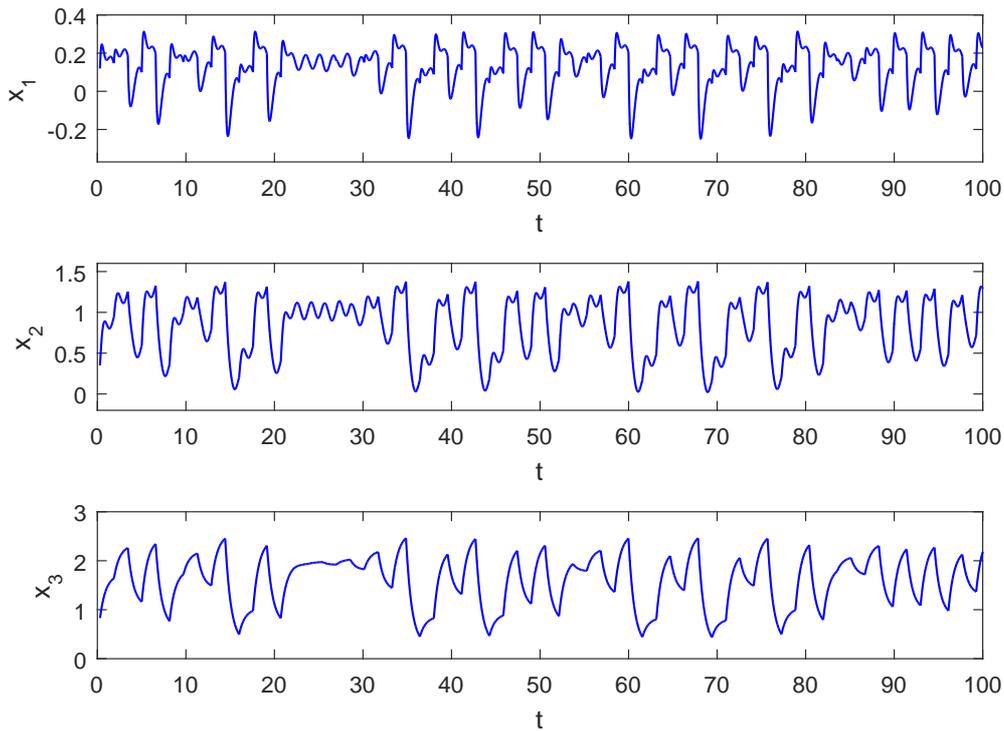}
\caption{Time-series of each coordinate of the solution of system (\ref{discontsyst3}) with initial data $x_1(0.3)= 0.12$, $x_2(0.3)= 0.35$, and $x_3(0.3)= 0.83$. The irregularity in each coordinate confirms the existence of an unpredictable solution.}
\label{quasifig2}
\end{figure}

In order to visualize the unpredictable behavior, we take an orbit $\left\{\widetilde{z}_k\right\}_{k \in \mathbb Z}$ of  the logistic map (\ref{logisticmap}) with $\lambda=3.91$ satisfying $\widetilde{z}_0=0.654$, and set up the system
\begin{eqnarray} \label{discontsyst3}
\begin{array}{l}
x_1'(t)=-5x_1(t)-2.5 x_2(t) + 0.1 \cos (x_3(t)) + \widetilde{g}_1(t) \\
x_2'(t)=2x_1(t)-3x_2(t) + 0.5 \sin (4t) + \widetilde{g}_2(t) \\
x_3'(t)=-1.5x_3(t) + 0.2 \tanh(x_1(t)) + \widetilde{g}_3(t),
\end{array}
\end{eqnarray} 
where $\widetilde{g}_1(t)= 4.5 \widetilde{z}_k$, $\widetilde{g}_2(t)= 3.7 \widetilde{z}_k$, and $\widetilde{g}_3(t)= 3.9 \widetilde{z}_k$ for $t \in [\theta_k, \theta_{k+1})$, $k \in \mathbb Z$.
Figure \ref{quasifig1} depicts the $3-$dimensional trajectory of system (\ref{discontsyst3}) corresponding to the initial data $x_1(0.3)= 0.12$, $x_2(0.3)= 0.35$, and $x_3(0.3)= 0.83$. Moreover, the time-series of each coordinate of the same trajectory is shown in Figure \ref{quasifig2} for $0.3 \leq t \leq 100$.  
Both Figure \ref{quasifig1} and Figure \ref{quasifig2} support the result of Theorem \ref{hypmainthm1} such that system (\ref{discontsyst2}) admits an exponentially asymptotically stable unpredictable solution. 

\section{Conclusion} \label{secconc}

The present results manifest that a discontinuous perturbation can make a system possess a continuous unpredictable solution. The existence and uniqueness of such a solution in quasilinear systems are rigorously proved using a Gronwall-Coppel type inequality. The smallness of the Lipschitz constant of the nonlinear term is required for the theoretical results. An unpredictable sequence is utilized in the perturbation function, which is the main source for the existence of an unpredictable solution, and the relation between their unpredictability constants is provided in the proof of Theorem \ref{hypmainthm1}. The result shows that they are proportional to each other.

The stability of the unpredictable solution is also discussed, and examples with unstable as well as exponentially asymptotically stable unpredictable solutions are provided. The advantage of asymptotic stability is the observation of unpredictability through simulations, in which the unpredictable behavior is displayed as irregularity. The main reason for the irregular behavior is the presence of Poincar\'e chaos in the dynamics where unpredictable functions are considered as points moving by shifts of the time argument \cite{Fen17a,Sell71}.


\end{document}